\documentclass{PoS}

\usepackage{graphicx,amsmath}
\usepackage[numbers]{natbib}

\title{Pulsar scattering in space and time}

\ShortTitle{Pulsar scattering}

\author{\speaker{Olaf Wucknitz}
  \thanks{This work is supported by the Emmy-Noether-Programme of the
    `Deutsche Forschungsgemeinschaft', reference WU\,588/1-1, and by 
    a Marie Curie European Reintegration Grant
    within the 7th European Community Framework Programme, Contract No.\
    PERG02-GA-2007-224897 `WIDEMAP'.}\\
        Max-Planck-Institut f\"ur Radioastronomie, Auf dem H\"ugel 69, 53121 Bonn, Germany,\\
  Argelander-Institut f\"ur Astronomie, Auf dem H\"ugel 71, 53121
  Bonn, Germany\\
        E-mail: \email{wucknitz@mpifr-bonn.mpg.de}}

\abstract{We report on a recent global VLBI experiment in which we study the scatter broadening of pulsars in the spatial and time domain simultaneously.
Depending on the distribution of scattering screen(s), geometry predicts that the less spatially broadened parts of the signal arrive earlier than the more broadened parts. This means that over one pulse period the size of the scattering disk should grow from pointlike to the maximum size. An equivalent description is that the pulse profile shows less temporal broadening on the longer baselines.
This contribution presents first results that are consistent with the
expected expanding rings. We also briefly discuss how the
autocorrelations can be used for amplitude calibration. This
requires a thorough investigation of the digitisation and the sampler
statistics and is not fully solved yet.
}

\FullConference{11th European VLBI Network Symposium \& Users Meeting,\\
		October 9--12, 2012\\
		Bordeaux, France}

\newcommand{\sub}[1]{_{\text{#1}}}
\newcommand{\rtext}[1]{\quad\text{#1}}

\begin{document}

\section{Introduction}
Any radiation we observe from pulsars or other sources has to pass the
interstellar medium (ISM), which can have a variety of effects. Most
important in our context is the fact that an ionised medium increases
the phase speed and reduces the group speed of the passing
radiation. As long as the density varies smoothly, this introduces
refraction. More important than this refraction are the effects of
density fluctuations in turbulent media. Small-scale
variations of the propagation speed lead to phase fluctuations,
producing local extrema of the light-travel time. According to
Fermat's theorem, sub-images (speckles) are formed at these local
extrema. Whenever there are many sub-images, they can be described as
being randomly distributed, with random phase differences. The
observer cannot resolve the individual images and sees the background
source scatter-broadened, provided that the intrinsic size is
sufficiently small. This is generally the case for pulsars, often for
maser sources, and sometimes for AGN cores.

\begin{figure}[hbt]
\centering\includegraphics[width=0.65\textwidth]{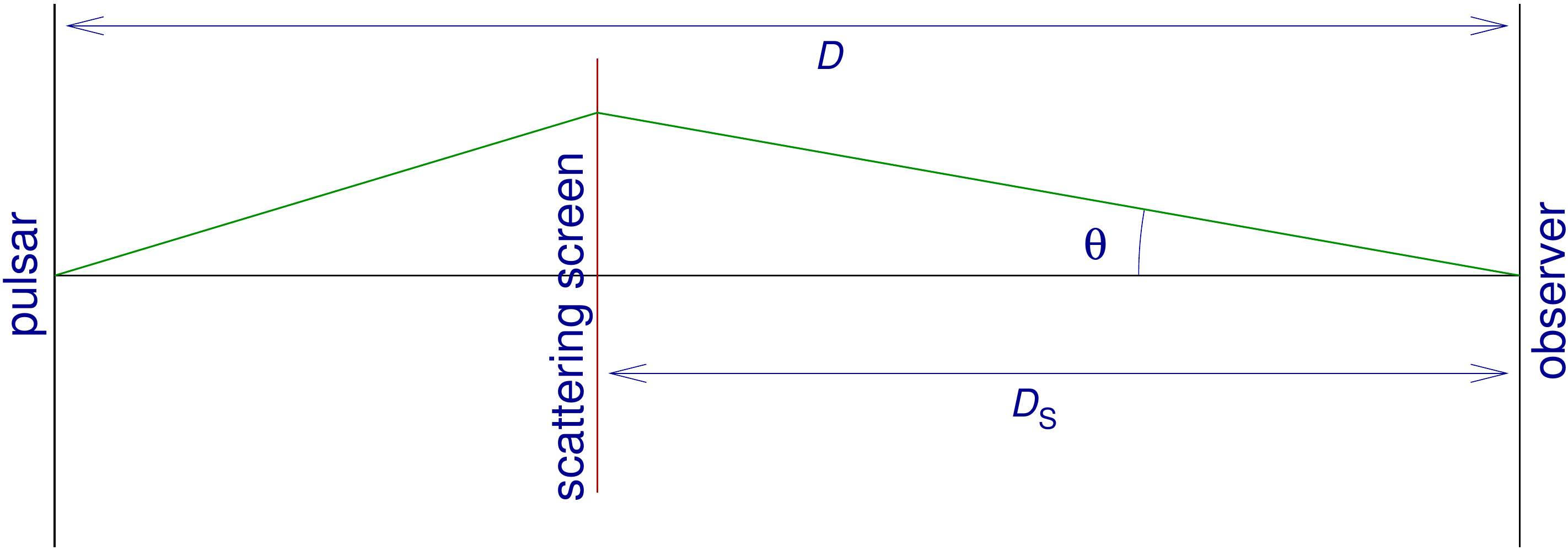}
\caption{Interstellar scattering geometry. A
  particular sub-image is deflected by an apparent angle $\theta$.}
\label{fig:geometry}
\end{figure}

As result of the effective deflection of the individual speckle, the
geometrical path is increased, which results in a scattering delay
(Fig.~\ref{fig:geometry}). This is similar to the situation in
gravitational lensing, with the difference that in lensing the Shapiro
(potential) delay contributes significantly to the total delay. In the
case of interstellar scattering, the direct delay caused by the ISM is
generally negligible.

For distances of the scattering medium and background source of $D\sub
s$ and $D$, the delay for small deflections $\theta$ can be derived
from the geometry in Fig.~\ref{fig:geometry} as
\begin{align}
\tau &= \frac12 \theta^2 D' \rtext{,}
&
D' &=\frac{D D\sub s}{D-D\sub s} \rtext{,}
\label{eq:tau D}
\end{align}
where the effective distance $D'$ is proportional to $D$ for a given
$D\sub s/D$. For a canonical situation in
which the scattering medium is in the middle between the background
source and the observer ($D\sub s=D/2$), we have $D'=D$.

Because different speckles have different deflections and delays, the
signal is broadened angularly and temporally.
The observed brightness depends on the relative phases of all
speckles, which become decorrelated over bandwidths larger than the reciprocal
scattering time. This scintillation can thus only be observed if the
spectral resolution is finer than $1/\tau$.

For Kolmogorov turbulence, we expect and generally observe a
wavelength dependence of 
\begin{align}
\theta\propto\lambda^{2.2} \rtext{,} \qquad\tau \propto \lambda^{4.4}
\rtext{.}
\end{align}

This conference contribution is about utilising the relation $\tau\propto
\theta^2$ not only for average scattering times $\tau$ and sizes
$\theta$, but also for individual speckles within one source.

\section{Scatter broadening in pulsars}

\begin{figure}[hbt]
\centering\includegraphics[width=0.45\textwidth]{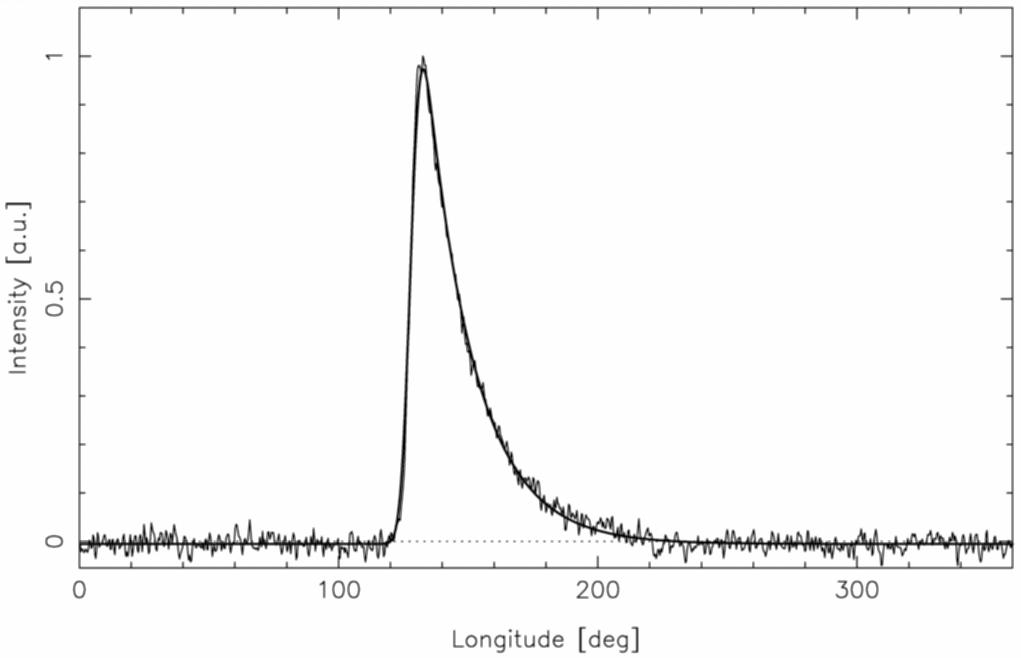}
\caption{Scatter-broadened profile of the pulsar B1815--15 at 1.4\,GHz
  \citep{loehmer2001}. The intrinsic pulse-width is
  sufficiently small to reveal the exponential scattering tail that
  is expected for a thin scattering screen.  This is also the target
  for which we show preliminary results in this contribution.}
\label{fig:profile}
\end{figure}

The intrinsic sizes of pulsars are much smaller than their scattering
sizes, so that the latter is not smeared out.
In our study we use pulsars because of the unique feature to offer
timing information in their radiation in the form of very regular and
often narrow pulses. It is this feature that allows us to measure the
scattering delays $\tau$ in a very direct way (Fig.~\ref{fig:profile}).

\section{Comparing angular and temporal broadening}

\begin{figure}[hbt]
\centering\includegraphics[width=0.4\textwidth]{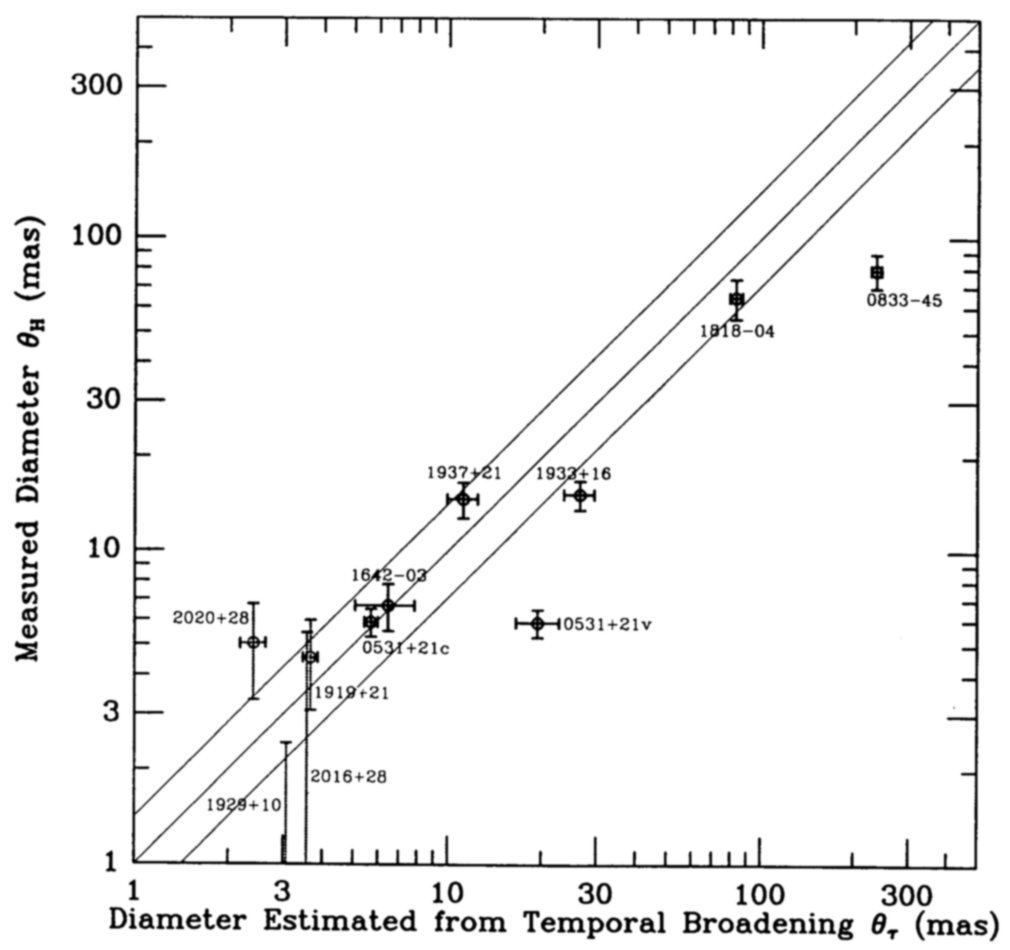}
\caption{Comparison of measured scattering size (vertical axis) with
  temporal broadening, shown horizontally as predicted scattering
  size as derived from $\tau$ via the equivalent of
  Eq.~\protect\eqref{eq:tau D} for a uniform distribution of
  scattering material \citep[observations at
    326\,MHz]{gwinn1993}. The diagonal band marks equality within
  errors.  }
\label{fig:gwinn}
\end{figure}

A comparison of characteristic scattering times $\tau$ and sizes
$\theta$ can be used to determine the mean distance of the
scattering screen(s) via Eq.~\eqref{eq:tau D}. Fig.~\ref{fig:gwinn}
shows this for a number of sources.
This comparison does already provide interesting information, but it
completely disregards the relation of $\tau$ vs.\ $\theta$ for different
speckles in the same system.
Because of this, it can, e.g., not distinguish between one scattering
screen or many scattering screens with the same mean distance.

\section{Secondary spectrum}

\begin{figure}[hbt]
\centering\includegraphics[width=0.65\textwidth]{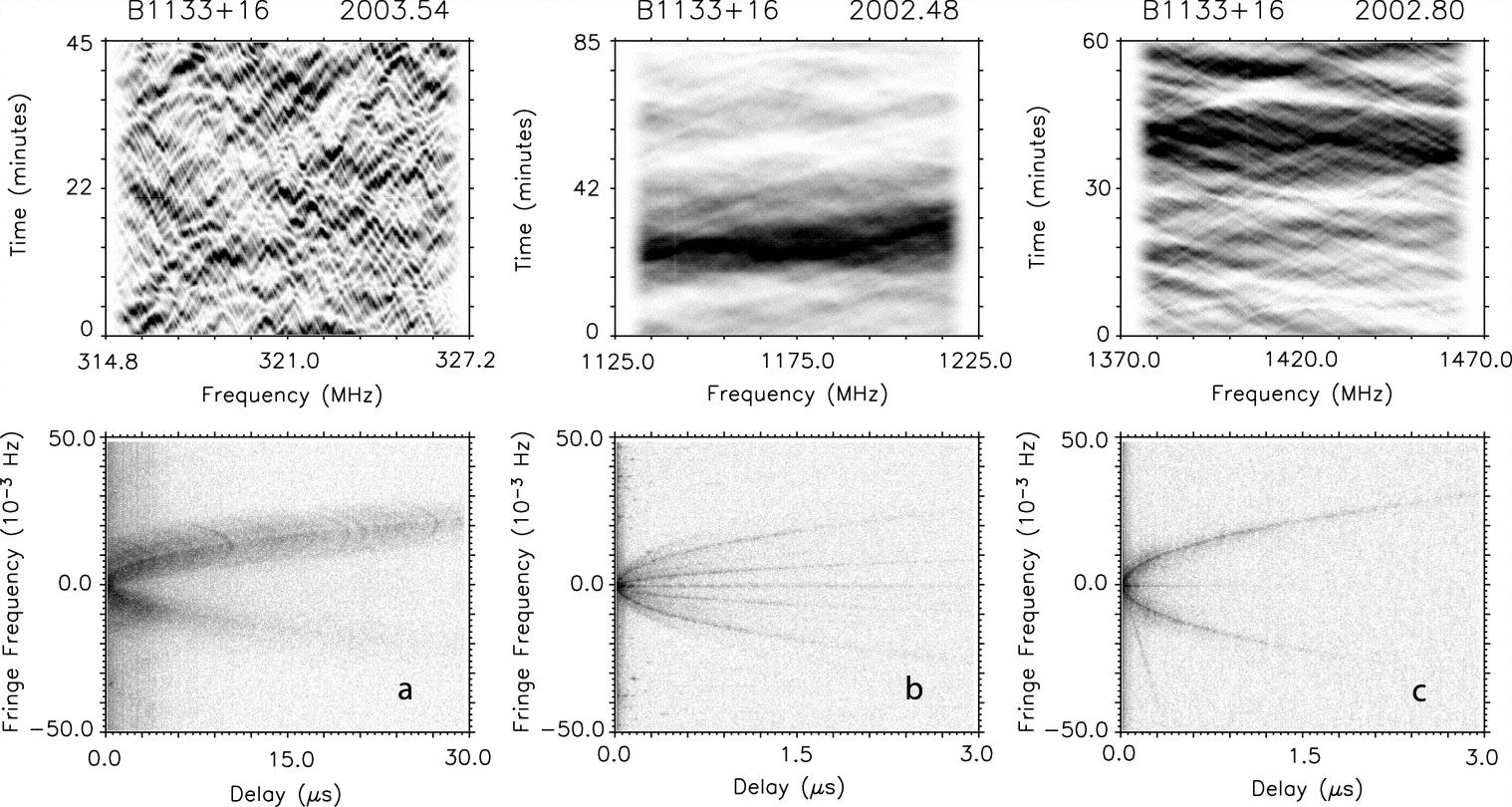}
\caption{Dynamic spectra (top) and secondary spectra (bottom) for
  three pulsars \citep{cordes2006}. The parabolas are an
  indirect result of the quadratic relation $\tau\propto\theta^2$.}
\label{fig:secondary}
\end{figure}

An alternative way of studying the relation between $\tau$ and
$\theta$ is provided by the secondary spectrum. To compute the
secondary spectrum, one starts with the power in a dynamic spectrum
(intensity as function of time and frequency) and then applies a
second two-dimensional Fourier transform that converts the frequency
axis into a delay axis and the time axis into a fringe-frequency axis
(Fig.~\ref{fig:secondary}).

The delay corresponds to delays between different speckles, while the
fringe-frequency measures the change with time due to proper motion
(equivalent to a projected Doppler effect).
The parabolic arcs (including the inverted arcs) that are generally
seen in the secondary spectra can be explained as resulting from
transverse motion (changing $\theta$) and the relation
$\tau\propto\theta^2$. This approach provides a wealth of information,
and can even be extended to interferometric observations
\citep{brisken2010}. However, it is model-dependent to some degree, because
it uses the Doppler effect as proxy for $\theta$ and is only sensitive
to a one-dimensional projection of $\theta$.

\section{Combined temporal and angular scattering study}

Neither of the methods discussed so far utilise all the available
information. The comparison of mean scattering times with sizes does
not use the full profiles, while the secondary spectrum does not use
the timing information provided by the pulsed emission. As one
possible way of combining the advantages of both methods, we do not
only want to study the temporal and spatial profiles independently but
measure the signal strength as function of $\tau$ and $\theta$, where
the deflection angle is even two-dimensional. The full
three-dimensional profile can potentially provide much more
information than the temporal and spatial profiles alone.

\begin{figure}[hbt]
\centering\includegraphics[width=0.69\textwidth]{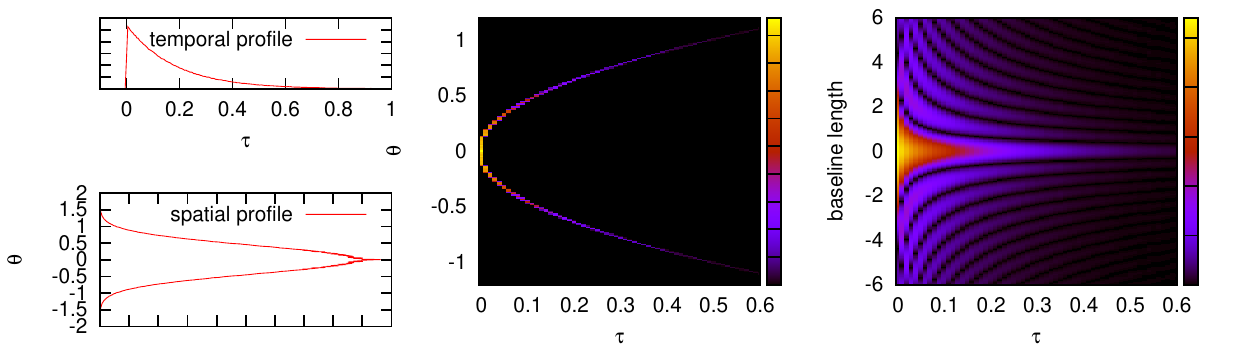} \\[1ex]
\includegraphics[width=0.69\textwidth]{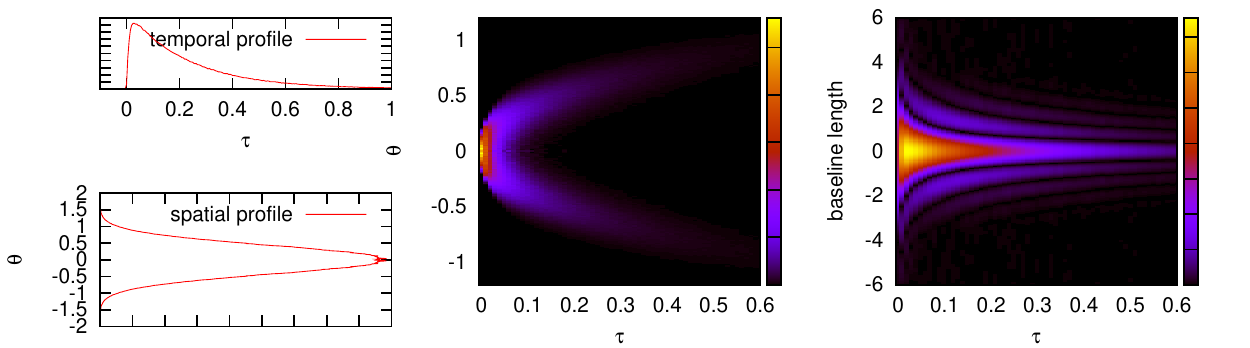} \\[1ex]
\includegraphics[width=0.69\textwidth]{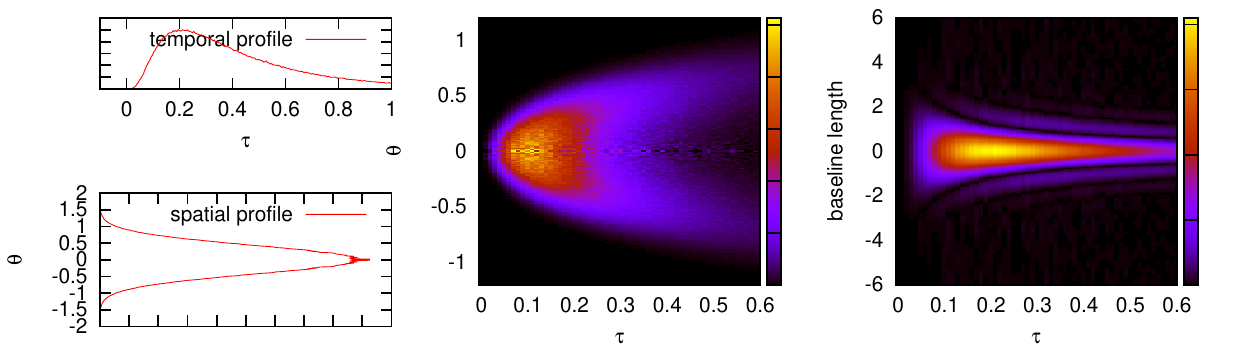}
\caption{Simulated profiles for one screen (top), two screens (middle)
and for a continuous scattering medium (bottom). Each panel shows the
temporal and spatial profiles, the combined profile as function of
$\tau$ and $\theta$ and the expected visibility amplitudes as function
of $\tau$ and baseline length.}
\label{fig:simul}
\end{figure}

Fig.~\ref{fig:simul} shows expected profiles as function of $\tau$,
$\theta$, and $\tau$ and $\theta$ combined, as well as expected
visibility amplitudes. We immediately see that a combined measurement
can distinguish between different distributions of scattering material
even if the temporal and spatial profiles are almost unchanged.

How can we measure the three-dimensional profile in practice? One
option is to produce interferometric images as function of scattering
delay $\tau$. In the case of a single scattering screen we expect to
see a point that grows into an expanding ring, the radius of which
goes with the square root of time (measured via apparent pulse phase).
For several scattering screens or a continuous medium we would see an
expanding fuzzy ring or blob instead.
Alternatively, and more appropriate for sparse uv coverage, we can
study the pulse profile as function of baseline length, where we would
expect to see less scattered profiles on longer baselines.

\section{Observations and analysis}

We decided to observe eight bright pulsars with global VLBI at
1.4\,GHz and 327/610\,MHz. The targets were selected according to
brightness and scattering parameters that allow us to resolve the
temporal and angular scattering. This project (GW022A/B) was observed
in June 2011 using the VLBA, five EVN stations and Arecibo. In the
following we only discuss very preliminary results for the pulsar
B1815--14 at 1.4\,GHz.  The baseband data were correlated in Bonn
using DiFX (with incoherent de-dispersion) running on computer
clusters at the MPIfR and AIfA. We first produced profiles from
autocorrelations in order to create matched filters. These filters
were then used for a gated correlation that was calibrated in
AIPS. Amplitudes were taken from unscattered control pulsars, while
the phases are determined from the targets. We then used these
solutions to calibrate a third correlation that was binned into 400
slots per period. Images have not been produced yet, but (complex)
profiles of calibrated visibilities.

\begin{figure}[hbt]
\centering\includegraphics[width=0.31\textwidth]{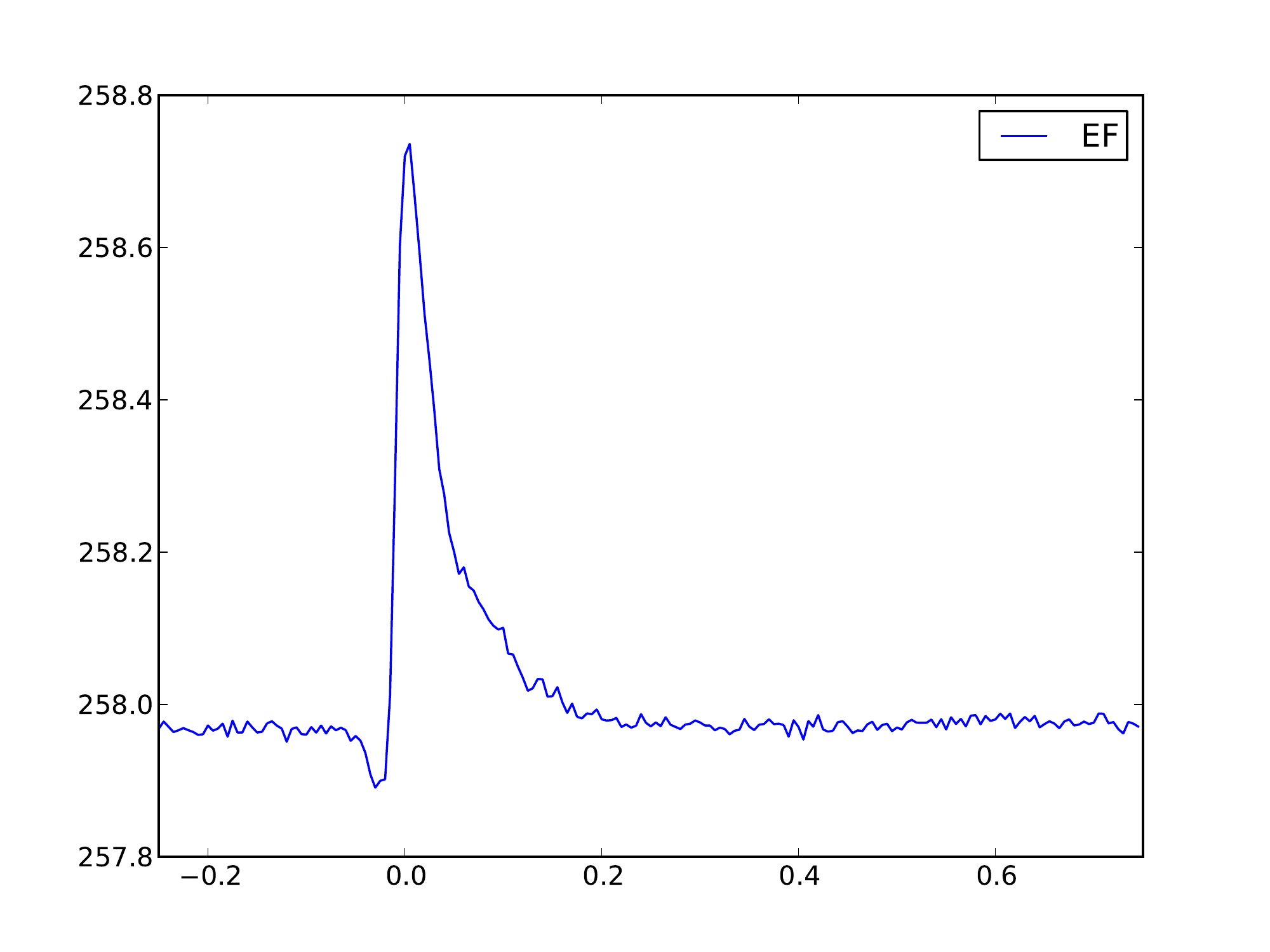}%
\includegraphics[width=0.31\textwidth]{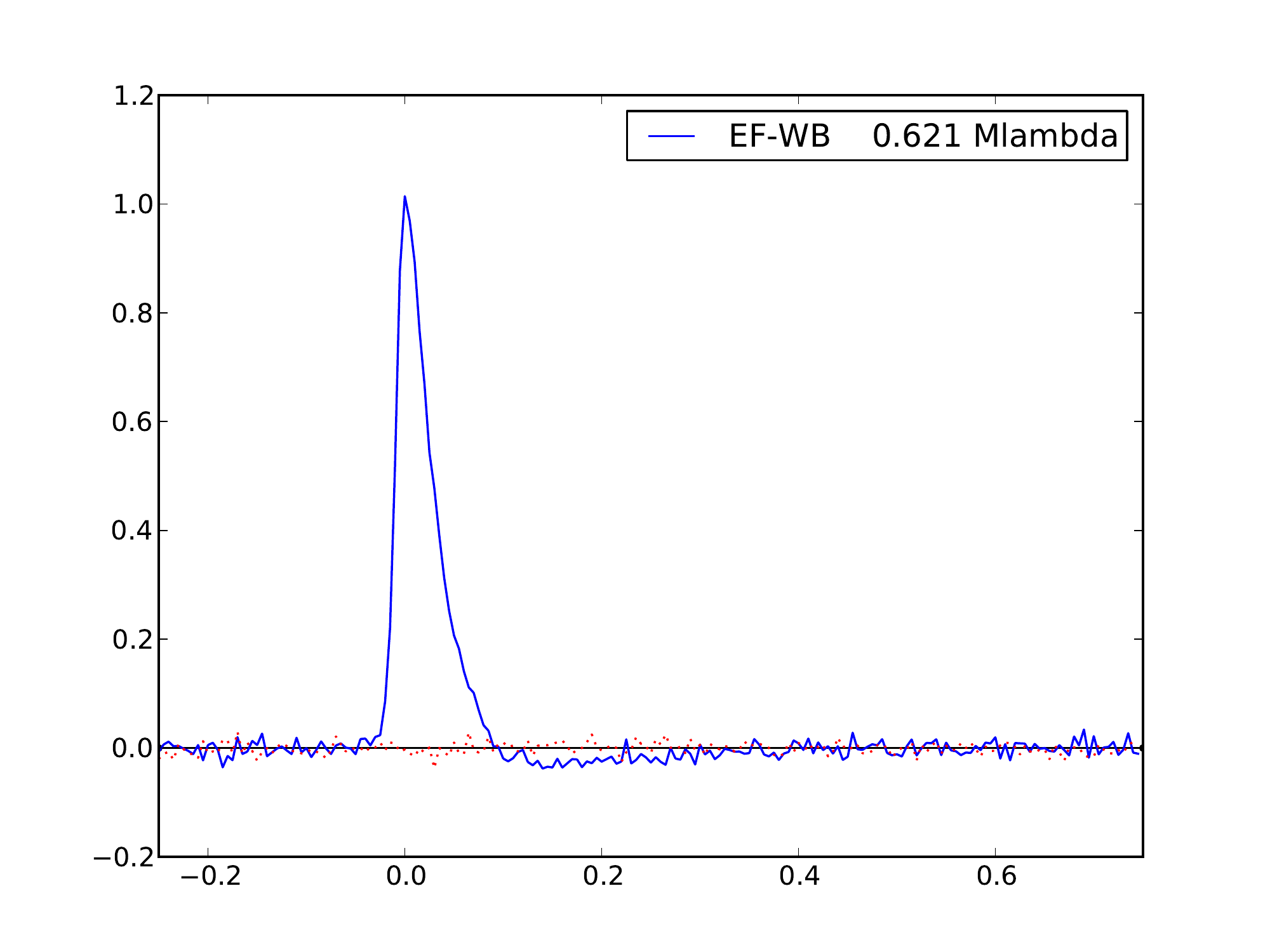}%
\includegraphics[width=0.31\textwidth]{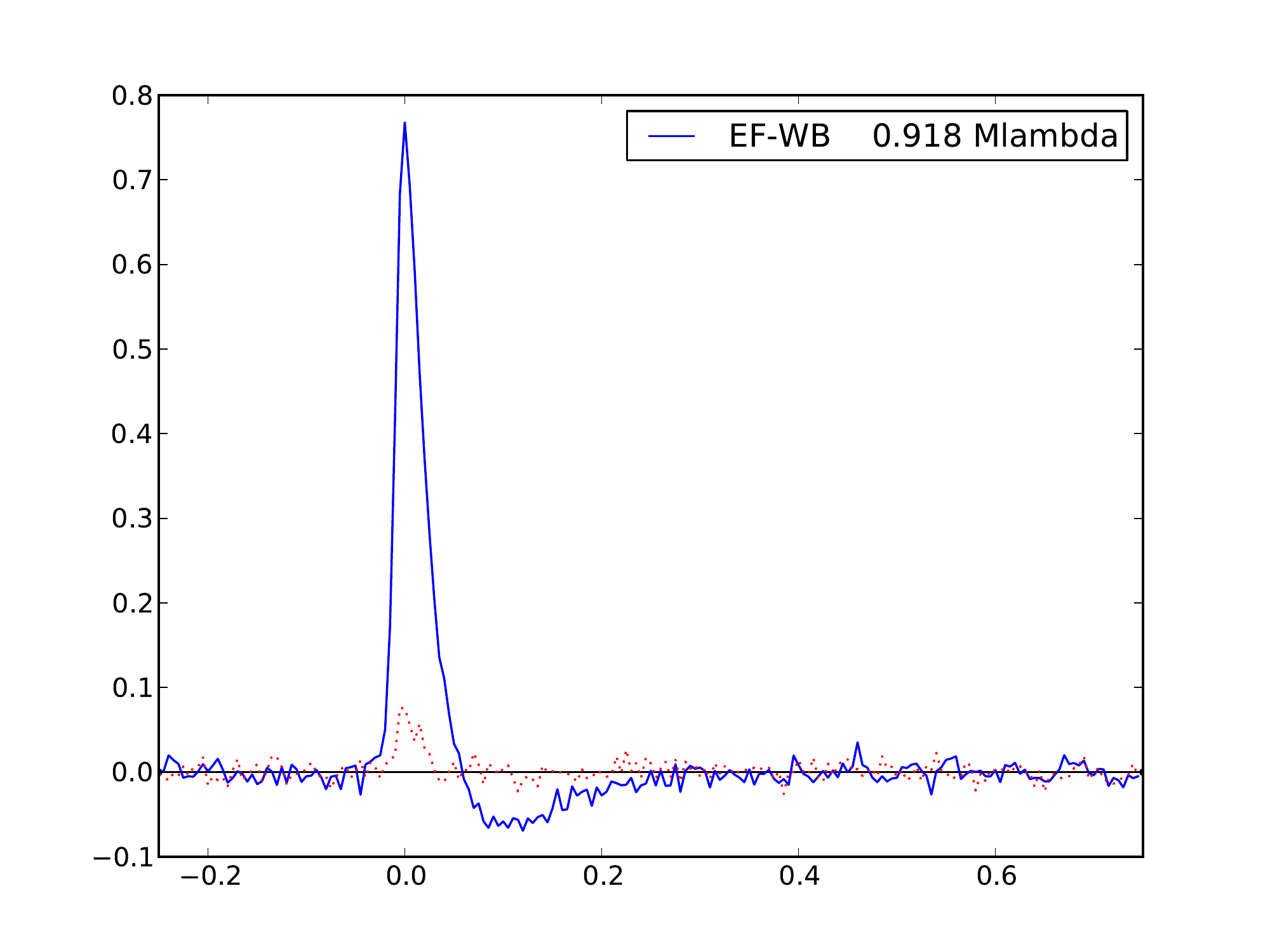}
\caption{Pulse profiles in autocorrelation (EF, left) and in two
  baselines (centre and right, 0.621\,k$\lambda$ and 0.918\,k$\lambda$) for one scan of about
  15\,min. The real part is shown in solid blue, the imaginary part in
  dotted red. The negative pre-pulse dip in the left plot results from
  digitisation non-linearities and dispersion.}
\label{fig:profiles}
\end{figure}

Fig.~\ref{fig:profiles} shows example profiles. We learn that the
profiles on longer baselines are indeed less scattered. In addition,
we already see that the angular scattering is inconsistent with
growing Gaussians, because the baseline profiles have negative
tails. These negative tails correspond to the negative
visibilities expected for expanding rings.

A full quantitative analysis that can constrain the distribution
of scattering material and measure its distance still has to be done.

\section{Outlook}

There is a chance to improve the essential amplitude calibration by
using information from the autocorrelations. In non-pulsar
observations this is not possible, because the
autocorrelations are generally dominated by system noise and not the
target signal. For folded pulsar data, the noise can be subtracted
easily, which opens this route for amplitude calibration. However, the
data were digitised with 2 bits per sample, which introduces
significant digitisation non-linearities. This can be taken into
account, provided that the sampler statistics are known. In our case
this is made more difficult by the switched power and (even more)
by the pulse-calibration signals that 
distort the statistics on a 1\,$\mu$sec period. This effect is most
significant for Arecibo where the occupation numbers are completely
skewed for certain pulse-cal period bins. Using an extended version of
DiFX we are able to derive amplitudes with about 10\,\% accuracy, but
this still has to be improved for the final calibration.

For some of the targets we will produce `movies' as function of 
scattering delay that may show the expected expanding
rings or blobs. The quantitative analysis will provide information
about the distribution and distance of the scattering material in
front of our target pulsars. Similar experiments are planned with LOFAR,
which samples a different part of parameter space.

\section{Acknowledgements}

The author wants to thank Joris Verbiest (MPIfR) for providing the pulsar ephemerides.

{\small\bibsep0.ex
\bibliographystyle{unsrtnat}
\bibliography{proc2}}

\begin{thebibliography}{4}
\providecommand{\natexlab}[1]{#1}
\providecommand{\url}[1]{\texttt{#1}}
\expandafter\ifx\csname urlstyle\endcsname\relax
  \providecommand{\doi}[1]{doi: #1}\else
  \providecommand{\doi}{doi: \begingroup \urlstyle{rm}\Url}\fi

\bibitem[{L{\"o}hmer} et~al.(2001){L{\"o}hmer}, {Kramer}, {Mitra}, {Lorimer},
  and {Lyne}]{loehmer2001}
O.~{L{\"o}hmer}, M.~{Kramer}, D.~{Mitra}, D.~R. {Lorimer}, and A.~G. {Lyne}.
\newblock \emph{\apjl}, 562:\penalty0 L157, 2001.

\bibitem[{Gwinn} et~al.(1993){Gwinn}, {Bartel}, and {Cordes}]{gwinn1993}
C.~R. {Gwinn}, N.~{Bartel}, and J.~M. {Cordes}.
\newblock \emph{\apj}, 410:\penalty0 673, 1993.

\bibitem[{Brisken} et~al.(2010)]{brisken2010}
W.~F. {Brisken} et~al.
\newblock \emph{\apj}, 708:\penalty0 232, 2010.

\bibitem[{Cordes} et~al.(2006){Cordes}, {Rickett}, {Stinebring}, and
  {Coles}]{cordes2006}
J.~M. {Cordes}, B.~J. {Rickett}, D.~R. {Stinebring}, and W.~A. {Coles}.
\newblock \emph{\apj}, 637:\penalty0 346, 2006.

\end{thebibliography}

\end{document}